\documentclass[twocolumn,eqsecnum,showkeys,byrevtex,secnumarabic,amssymb,
nobibnotes,aps,pra]{revtex4-1}
\usepackage{dsfont}
\usepackage{amsfonts}
\usepackage{amssymb,latexsym,amsmath,amsthm}
\usepackage{array}
\usepackage{amscd}
\usepackage{graphicx}
\usepackage{xy}
\xyoption{all}

\newcommand{\rme}{\textit{e}}     %
\newcommand{\rmd}{\textit{d}}     %
\newcommand{\rmi}{\textit{i}}     %
\renewcommand{\Re}{\mathfrak{Re}} %
\renewcommand{\Im}{\mathfrak{Im}} %

\begin{document}

\title{Complementarity versus coordinate transformations:\\ mapping between pseudo-Hermiticity and weak pseudo-Hermiticity}\thanks{This paper is dedicated to the memory of Pr. Mustapha Bentaiba, beloved friend and esteemed colleague and teacher.}

\author{Samira Saidani and Sid-Ahmed Yahiaoui}%
\email[Corresponding author: ]{\tt{s$\_$yahiaoui@univ-blida.dz}}
\affiliation{D\'epartement de physique, facult\'e des sciences, universit\'e Sa\^ad DAHLAB-Blida~1, B.P. 270 Route de Soum\^aa,
09000 Blida, Algeria}
\date{\today}

\begin{abstract}
\noindent We study the concept of the complementarity, introduced by Bagchi and Quesne in [Phys. Lett. A {\bf 301}, 173 (2002)], between pseudo-Hermiticity and weak pseudo-Hermiticity in a rigorous mathematical viewpoint of coordinate transformations when a system has a position-dependent mass. We first determine, under the modified-momentum, the generating functions identifying the complexified potentials $V_\pm(x)$ under both concepts of pseudo-Hermiticity $\widetilde\eta_+$ (resp. weak pseudo-Hermiticity $\widetilde\eta_-$). We show that the concept of complementarity can be understood and interpreted as a coordinate transformation through their respective generating functions. As consequence, a similarity transformation which implements coordinate transformations is obtained. We show that the similarity transformation is set up as fundamental relationship connecting both $\widetilde\eta_+$ and $\widetilde\eta_-$. A special factorization $\eta_+=\eta_-^\dagger \eta_-$ is discussed in the case of a constant mass and some B\"acklund transformations are derived.
\end{abstract}
\pacs{xx;yy;zz}

\maketitle

\section{Introduction and preliminaries}%

\noindent The basic mathematical structure underlying the properties of pseudo-Hermiticity was first revealed in Mostafazadeh's papers series \cite{1}. This subject has attracted much attention of physicists and many papers have been written and reviewed \cite{2,3,4,5,6,7,8,9,10,11,12} (Ref.~\cite{2} contains a detailed bibliography and list of vast references) and it has been well established that the concept of pseudo-Hermiticity is more general then those of Hermiticity and $\mathcal {PT}$-symmetry \cite{13,14,15,16,17,18}.\\
\indent By definition, a linear operator (e.g. Hamiltonian) $H:\mathfrak{H}\rightarrow\mathfrak{H}$ acting in a Hilbert space $\mathfrak{H}$ is called $\eta$-pseudo-Hermitian if it obeys to \cite{1,2,3,4}:
\begin{eqnarray}\label{1.1}
  H^\dagger = \eta H \eta^{-1},
\end{eqnarray}
where $\eta:\mathfrak{H}\rightarrow\mathfrak{H}$ is a Hermitian, linear and invertible operator and $H^\dagger$ denotes the Hermitian adjoint of $H$. Then (non-Hermitian) Hamiltonian $H$ has a real-spectrum \cite{2,13} if there is an invertible, linear and a Hermitian operator $\zeta:\mathfrak{H}\rightarrow\mathfrak{H}$, such that $\eta=\zeta^\dagger\zeta$ is hermitian operator. In these settings, the reality of the bound-state eigenvalues of $H$ can be associated with $\eta$-pseudo-Hermiticity. Note that setting $\eta=\mathds 1$ reduces the assumption \eqref{1.1} to the usual well-known concept of Hermiticity, while $\eta=\mathcal{PT}$ implies that a Hamiltonian is $\mathcal{PT}$-symmetric, i.e. $H=H^{\mathcal {PT}}$, see Ref.~\cite{13}.
\par Soon after the introduction of the concept of pseudo-Hermiticity, Solombrino \cite{6} introduced the notion of weak pseudo-Hermiticity in order to report the existence, at least for {\it diagonalizable operators}, of a large class of operators satisfying Eq.~\eqref{1.1} without any constraint on $\eta$ and coinciding with the class of all pseudo-Hermitian operators. Subsequently Bagchi and Quesne \cite{7} used the term {\it complementarity} in order to describe the relationship connecting the twin concepts of pseudo-Hermiticity and weak pseudo-Hermiticity. They point out that it is possible to break up $\eta$ into two operators, i.e. $\eta_+$ and $\eta_-$, following combinations:
\begin{equation}\label{1.2}
  \eta_+ H = H^\dagger\eta_+
  \qquad {\rm and} \quad
  \qquad \eta_- H = H^\dagger\eta_-,
\end{equation}
where $\eta_\pm=\eta\pm\eta^\dagger$. The first assumption in \eqref{1.2} corresponds to the pseudo-Hermiticity with a second-order differential realization of $\eta_+$, while the second is associated with a weak pseudo-Hermiticity and corresponds to a first-order realization. Such differential realizations for the operators $\eta_\pm$ are possible because $\eta_+$ is Hermitian ($\eta_+^\dagger=\eta_+$), while $\eta_-$ is anti-Hermitian ($\eta_-^\dagger=-\eta_-$). Later, Mostafazadeh has suggested "{\it a careful reexamination of the equivalence of these concepts}" for a large class of linear operators {\it not} necessarily diagonalizable \cite{4}.\\
\indent In this paper, we would return to the complementarity principle describing the relationship connecting both twin concepts of pseudo-Hermiticity and weak pseudo-Hermiticity. In their work, Ref.~\cite{7}, Bagchi and Quesne suggest "{\it that weak pseudo-Hermiticity is not more general than pseudo-Hermiticity but works complementary to it}". However the term {\it complementarity} therein was neither explicitly stated nor its mathematical argumentation was given, so we think that more general interpretation is yet to be formulated. Besides the proposal made by Mostafazadeh, in Ref.~\cite{4}, which consists in establishing, algebraically, the equivalence between these concepts for a large class of linear operators, we want to tackle the concept of complementarity otherwise in the framework of position-dependent mass (PDM), since we believe that PDM allows us to better understand the case of the usual constant mass ; this latter can be deduced from the PDM background by setting $m(x)=1$. We aim in this paper to fill this gap and to point out that the concept of complementarity can be understood from a rigorous mathematical viewpoint and interpreted as a coordinate transformation, implemented by a similarity transformation, leading to connect the twin concepts of pseudo-Hermiticity and weak pseudo-Hermiticity.
\par We want to point out that we preferred used the term {\it mapping} raised in the title of our paper instead of the term {\it equivalence} used by Mostafazadeh in Ref.~\cite{4}, since our main working tool in this paper is about coordinate transformations. Before performing these transformations, we purpose to establish, in PDM background, the following results which constitute the body of our paper: we start in Sec. II by formulating and deducing the mathematical structures underlying the properties of pseudo-Hermiticity and derive the general expression of the continuity equation in the PDM background under a modified-momentum $p\mapsto\widehat\pi=p-A(x)/U(x)$. In Sec.~III we generate the functions that lead to identify the complexified potentials $V_+(x)$ (resp. $V_-(x)$) under pseudo-Hermiticity (resp. weak pseudo-Hermiticity) $\widetilde\eta_+$ (resp. $\widetilde\eta_-$). In Sec.~IV we show that both deduced generating functions are connected through coordinate transformations, hence establishing the connection between the complementarity and coordinate transformation. This connection takes us to find a mapping relating $\widetilde\eta_+$ to $\widetilde\eta_-$ through some similarity transformation, which implements the coordinate transformations. We consider the case of a constant mass and the special factorization $\eta_+=\eta_-^\dagger\eta_-$ is discussed, as well as its mathematical implications. Finally we present our conclusions in Sec.~V.

\section{Pseudo-Hermiticity and position-dependent mass}%

\noindent Before we begin by addressing the different points mentioned hereinabove, let us establish some mathematical properties underlying the pseudo-Hermiticity within PDM background. To this end we consider a von Roos' one-dimensional spatially varying mass Hamiltonian \cite{19}:
\begin{eqnarray*}
H&=&\dfrac{1}{4}\left(M^\alpha(x)pM^\beta(x)pM^\gamma(x)+M^\gamma(x)pM^\beta(x)pM^\alpha(x)\right)\nonumber \\
&&+V(x),
\end{eqnarray*}
which has an advantage to keep $H$ Hermitian. $M(x)=m_0m(x)$ and $\alpha,\,\beta$ and $\gamma$ are three parameters obeying to the restriction $\alpha+\beta+\gamma=-1$ in order to grant the classical limit and $V:\mathbb R\rightarrow\mathbb C,\,x\mapsto V(x)=V_{\rm Re}(x)+\rmi V_{\rm Im}(x)$ is complex-valued potential. Here $p\big(\equiv-\rmi\rmd/\rmd x\big)$ is a momentum with $\hbar=m_0=1$ and $m(x)$ is some dimensionless-real valued mass.\\
\indent Using the ordering prescription adopted by BenDaniel-Duke (i.e. $\alpha=\gamma=0$ and $\beta=-1$) \cite{20}, then the last Hamiltonian becomes:
\begin{equation}\label{3}
  H = p\,U^2(x) p+V(x),
\end{equation}
where $U:\mathbb R\rightarrow\mathbb R$ is real-valued function defined by $U(x)=1/\sqrt{2m(x)}$. Operating a shift by replacing the momentum  $p$ with the canonical momentum
 $\hat\pi$, such that
\begin{equation*}
p\quad\mapsto\quad \hat\pi=p-\frac{A(x)}{U(x)},
\end{equation*}
where $A:\mathbb R\rightarrow\mathbb C,\,x\mapsto A(x)=a(x)+\rmi\, b(x)$ is complex-valued function, we can show that the new modified-Hamiltonian,
\begin{equation}\label{4}
H\quad\mapsto\quad \mathcal H = \left(p-\frac{A(x)}{U(x)}\right)U^2(x)\left(p-\frac{A(x)}{U(x)}\right)+V(x),
\end{equation}
is pseudo-Hermitian in the sense improved by Mostafazadeh \cite{1,2,3}. In the following $\mathcal P$ and $\mathcal T$ denote parity and time-reversal operators, respectively.
\par Let us consider an antilinear and invertible operator $\widetilde\tau$ satisfying: $\widetilde\tau=\mathcal T\rme^{\rmi\alpha(x)}$ and $\mathcal H^\dagger=\widetilde\tau \mathcal H\widetilde\tau^{-1}$, where:
\begin{equation}\label{qq}
 \alpha(x)=-2\int^x A(x')\,\rmd\mu(x'),
\end{equation} 
is some complex-valued function with respect of the measure $\rmd\mu(x)=U^{-1}(x)\rmd x$. Then it is easy to check that $\widetilde\tau$ is Hermitian, i.e. $\widetilde\tau=\widetilde\tau^\dagger$, and the use of the gauge-like transformation:
\[
\rme^{\mp\rmi\alpha(x)}p\,\rme^{\pm\rmi\alpha(x)}=p\pm\frac{\rmd\alpha(x)}{\rmd x},\qquad
\rme^{\mp\rmi\alpha(x)}x\,\rme^{\pm\rmi\alpha(x)}=x,
\]
led us to establish the anti-pseudo-Hermiticity of the modified Hamiltonian \eqref{4} with respect to $\widetilde\tau$, namely
\begin{widetext}
\begin{eqnarray}\label{5}
  \widetilde\tau \mathcal H\widetilde\tau^{-1} &=& \mathcal T \rme^{\rmi\alpha(x)}\bigg(p-\frac{A(x)}{U(x)}\bigg)U^2(x)\bigg(p-\frac{A(x)}{U(x)}\bigg)
                               \rme^{-\rmi\alpha(x)}\mathcal T+ \mathcal T \rme^{\rmi\alpha(x)}V(x)\rme^{-\rmi\alpha(x)}\mathcal T\nonumber  \\
    &=& \mathcal T\bigg(p-\frac{A(x)}{U(x)}-\frac{\rmd\alpha(x)}{\rmd x}\bigg)U^2(x)\bigg(p-\frac{A(x)}{U(x)}-\frac{\rmd\alpha(x)}{\rmd x}\bigg) \mathcal T + \mathcal T V^\ast(x)\mathcal T\nonumber \\
    &=& \mathcal T\bigg(p+\frac{A(x)}{U(x)}\bigg)U^2(x)\bigg(p-\frac{A(x)}{U(x)}\bigg) \mathcal T + \mathcal T V^\ast(x)\mathcal T\nonumber \\
    &=& \mathcal T\bigg(p-\frac{A^\ast(x)}{U(x)}\bigg)U^2(x)\bigg(p-\frac{A^\ast(x)}{U(x)}\bigg) \mathcal T + \mathcal T V^\ast(x)\mathcal T\nonumber \\
    &\equiv& \mathcal H^\dagger,
\end{eqnarray}
\end{widetext}
where $\mathcal T^\dagger = \mathcal T$ and $\mathcal T f(x,p)\mathcal T = f^\ast(x,-p)$ for every function $f:\mathbb R\rightarrow\mathbb C$.
\par Further it may be shown that for every $\mathcal{PT}$-symmetric Hamiltonian $\mathcal H$ given by Eq.~\eqref{4}, there exists a linear and invertible operator $\widetilde\eta=\exp\big\{-\rmi\,\alpha(x)\big\}\mathcal P$ such that:
\begin{eqnarray}\label{2.4}
  \widetilde\eta^\dagger &=& \mathcal P \exp\left[-2\rmi\int^x \frac{A^\ast(x')}{U(x')}\,\rmd x'\right] \nonumber \\
               &=& \exp\left[2\rmi\int^x \frac{A^\ast(-x')}{U(-x')}\,\rmd x'\right] \mathcal P \nonumber\\
               &=& \exp\left[2\rmi\int^x \frac{\Re A(x')-\rmi\,\Im A(x')}{U(x')}\,\rmd x'\right] \mathcal P \nonumber \\
               &=& \exp\left[2\rmi\int^x \frac{A(x')}{U(x')}\,\rmd x'\right]\mathcal P \nonumber \\
               &\equiv& \widetilde\eta,
\end{eqnarray}
is an Hermitian operator with respect of the measure $\rmd\mu(x)$, where we used $\mathcal P^\dagger = \mathcal P$ and $\mathcal P f(x,p)\mathcal P = f(-x,-p)$, for every function $f:\mathbb R\rightarrow\mathbb C$. In Eq.~\eqref{2.4}, we demand to the real and imaginary parts of $A(x)$ to be even and odd functions, respectively, and $U(-x)=U(x)$. Then the product of $\mathcal{PT}$-symmetry and $\widetilde\tau$-anti-pseudo-Hermitian operator coincides with $\widetilde\eta$-pseudo Hermiticity, i.e. $\widetilde\eta \equiv \widetilde\tau\mathcal{PT}$ \cite{3} and adapts very well to the modified-Hamiltonian \eqref{4} associated to quantum systems endowed with position-dependent mass.\\
\indent Now let us derive the general expression of the continuity equation for class of the modified-Hamiltonian $\mathcal H$ given by Eq.~\eqref{4}. Then the one-dimensional PDM Schr\"odinger equation and its $\mathcal{PT}$-symmetric conjugate read as:
\begin{widetext}
\begin{eqnarray}
  \rmi\frac{\partial}{\partial t}\Psi(x,t) &=& \left(\rmi\frac{\partial}{\partial x}+\frac{A(x)}{U(x)}\right)U^2(x)\left(\rmi\frac{\partial}{\partial x}+\frac{A(x)}{U(x)}\right)\Psi(x,t)+V(x)\Psi(x,t), \label{2.5}\\
  -\rmi\frac{\partial}{\partial t}\Psi^\ast(-x,t) &=& \left(\rmi\frac{\partial}{\partial x}+\frac{A^\ast(x)}{U(x)}\right)U^2(x)\left(\rmi\frac{\partial}{\partial x}+\frac{A^\ast(x)}{U(x)}\right)\Psi^\ast(-x,t)+V(x)\Psi^\ast(-x,t),\label{2.6}
\end{eqnarray}
\end{widetext}
where $\mathcal{PT} V(x)=V^\ast(-x)\equiv V(x)$ and $\mathcal{PT}\frac{\partial}{\partial t}\Psi(x,t)=\frac{\partial}{\partial t}\Psi^\ast(-x,t)$.
\par Following \cite{1,7}, consider that $\Psi_1(x,t)$ and $\Psi^\ast_2(-x,t)$ satisfy both of Eqs.~\eqref{2.5} and \eqref{2.6}, respectively. Next using Eq.~\eqref{2.4} and multiplying Eq.~\eqref{2.5} by $\widetilde\eta^\dagger\,\Psi^\ast_2(-x,t)$ and Eq.~\eqref{2.6} by $\widetilde\eta\,\Psi_1(x,t)$, the subtraction of the both resulting equations yields:
\begin{widetext}
\begin{eqnarray}\label{2.7}
\rmi\frac{\partial}{\partial t}\,\Psi^\ast_2(-x,t)\,\widetilde\eta\,\Psi_1(x,t) &=& -\frac{\partial}{\partial x}\,\widetilde\eta\, U^2(x)\left(\Psi^\ast_2(-x,t)\frac{\partial}{\partial x}\Psi_1(x,t)-\Psi_1(x,t)\frac{\partial}{\partial x}\Psi^\ast_2(-x,t)\right) \nonumber \\
  &&+\widetilde\eta\left[A^2(x)-A^{\ast2}(x)+\rmi \frac{d}{dx}U(x)\left(A(x)-A^\ast(x)\right)\right]\Psi_1(x,t)\Psi^\ast_2(-x,t) \nonumber \\
  &&+2\,\rmi\, \widetilde\eta \,U(x)\Big(A(x)-A^\ast(x)\Big)\Psi^\ast_2(-x,t)\frac{\partial}{\partial x}\Psi_1(x,t).
\end{eqnarray}
\end{widetext}

\indent In order to reduce Eq.~\eqref{2.7} to a continuity equation for a quantum system endowed with PDM in its traditional form, i.e.
\begin{eqnarray}\label{2.8}
  \frac{\partial}{\partial t}\,\rho_{\widetilde\eta}(x,t)+\frac{\partial}{\partial x}\,J_{\widetilde\eta}(x,t) = 0,
\end{eqnarray}
where $\rho_{\widetilde\eta}(x,t)$ and $J_{\widetilde\eta}(x,t)$ are given through:
\begin{eqnarray}
  \rho_{\widetilde\eta}(x,t) &=&  \Psi^\ast_2(-x,t)\,\widetilde\eta\,\Psi_1(x,t), \nonumber\\
  J_{\widetilde\eta}(x,t)    &=& -\rmi\,\widetilde\eta\, U^2(x)\bigg[\Psi^\ast_2(-x,t)\frac{\partial}{\partial x}\Psi_1(x,t)\nonumber \\
  &&-\Psi_1(x,t)\frac{\partial}{\partial x}\Psi^\ast_2(-x,t)\bigg],\nonumber
\end{eqnarray}
we must impose the following constraint: $A(x)=A^\ast(x)$, i.e. $A(x)\equiv a(x)\in\mathbb R$ and $b(x)=0$. Indeed, these constraints agree with those deduced in Refs.~\cite{5,7} and will be obtained otherwise in Sec.~III and used later throughout this article. By proceeding to integrate Eq.~\eqref{2.8} over the entire real line, taking into account $\Psi_{1,2}(\pm\infty,t)=0$ and the fact that $\widetilde\eta$ does not depend on the temporal variable $t$, we deduce the conservation law, i.e.
\begin{eqnarray}\label{2.9}
   \frac{\partial}{\partial t}\int_{\mathbb R}\Psi^\ast_2(-x,t)\,\widetilde\eta\,\Psi_1(x,t)\,\rmd x = 0.
\end{eqnarray}
\indent In the framework of PDM, we can express $\Psi_{1,2}(x,t)$ in their respective energy representation:
\begin{eqnarray*}
    \Psi_1(x,t) &\sim & \frac{\psi_1(x)}{\sqrt{U(x)}}\,\rme^{-\rmi\, U(x)E_1t},\nonumber \\
    \Psi^\ast_2(-x,t) &\sim & \frac{\psi^\ast_2(-x)}{\sqrt{U(x)}}\,\rme^{\rmi\, U(x)E^\ast_2t},
\end{eqnarray*}
then Eq.~\eqref{2.9} becomes:
\begin{eqnarray}\label{2.10}
   (E_1-E^\ast_2)\int_{\mathbb R}\psi^\ast_2(-x)\,\widetilde\eta\,\psi_1(x)\,\rmd x = 0,
\end{eqnarray}
which is well-known as the $\widetilde\eta$-orthogonality condition (see, Refs.~\cite{1,7}). Of course the condition \eqref{2.10} holds if, and only if, it satisfies either the reality of the energy spectrum (i.e. $E_1=E^\ast_2$) and/or the condition $\widetilde\eta\,\psi_1(x)=0$. Once again, both constraints will be deduced and used in Sec.~III.\\
\indent Obviously the condition \eqref{2.10} can be transformed into the $\mathcal{PT}$-orthogonality condition \cite{5,7}, i.e.
\begin{eqnarray}\label{2.11}
   (E_1-E^\ast_2)\int_{\mathbb R}\psi_2^{\mathcal{PT}}(x)\,\psi_1(x)\,\rmd x = 0,
\end{eqnarray}
if $A(x)=0$ and $\psi_2^{\mathcal{PT}}(x)=\psi^\ast_2(-x)$, since by invoking the following gauge transformations: $\psi_1(x)\mapsto\sqrt{\widetilde\eta}\,\psi_1(x)$ and $\psi^\ast_2(-x)\mapsto\sqrt{\widetilde\eta}\,\psi^\ast_2(-x)$, as defined in Eq.~\eqref{2.4}, allow us to derive Eq.~\eqref{2.10} from Eq.~\eqref{2.11}.

\section{Generating functions and their associated potentials}%

\noindent Let us start by decomposing $\eta_+\left(\equiv \zeta^\dagger \zeta\right)$
following the supersymmetry of quantum mechanics, where the operators $\zeta$ and $\zeta^\dagger$ are obtained with the first-order differential realization:
\begin{equation}\label{7}
  \zeta = U(x)\frac{\rmd}{\rmd x}+W(x), \quad \zeta^\dagger = -\frac{\rmd}{\rmd x}\,U(x)+W^\ast(x),
\end{equation}
where $W(x)=F(x)+\rmi\, G(x)$, and $F(x)$ and $G(x)$ are some real functions to be determined.
\par Under the shift on the momentum $p\mapsto \hat\pi=p-A(x)/U(x)$, both operators in Eq.~\eqref{7} can be recast in their new forms:
\begin{equation}\label{8}
\begin{split}
  \zeta\quad&\mapsto\quad\widetilde\zeta = U(x)\frac{\rmd}{\rmd x}-\rmi A(x)+W(x),\\
  \zeta^\dagger\quad&\mapsto\quad\widetilde\zeta^\dagger = -\frac{\rmd}{\rmd x}\,U(x)+\rmi A^\ast(x)+W^\ast(x),
\end{split}
\end{equation}
in such a way that $\eta_+$ is transformed into $\widetilde\eta_+\left(\equiv \widetilde\zeta^\dagger \widetilde\zeta\right)$ as:
\begin{equation}\label{9}
  \eta_+\quad\mapsto\quad\widetilde\eta_+ = -U^2(x)\frac{\rmd^2}{\rmd x^2}-2\mathcal K(x)\frac{\rmd}{\rmd x}+\mathcal L(x),
\end{equation}
where the functions $\mathcal K(x)$ and $\mathcal L(x)$ are defined by:
\begin{eqnarray*}
  \mathcal K(x) &=& U(x)U'(x)-\rmi\, U(x)(a(x)-G(x)), \\
  \mathcal L(x) &=& (b(x)+F(x))^2+(a(x)-G(x))^2 \\
  &+&\rmi\frac{\rmd}{\rmd x}\Big\{U(x)\Big[a(x)+\rmi\Big(b(x)+F(x)+\rmi\, G(x)\Big)\Big]\Big\},
\end{eqnarray*}
and the prime denotes differentiation with respect to the variable $x$. It is easy for the reader to check that $\widetilde\eta_+$ is indeed Hermitian, i.e. $\widetilde\eta^\dagger_+=\widetilde\eta_+$.
\par On the other hand, the modified-Hamiltonian \eqref{4} can be expressed as:
\begin{equation}\label{10}
  \mathcal H = -U^2(x)\frac{\rmd^2}{\rmd x^2}-2\mathcal M(x)\frac{\rmd}{\rmd x}+\mathcal N(x)+V_+(x),
\end{equation}
where the functions $\mathcal M(x)$ and $\mathcal N(x)$ are defined in terms of the real-functions $a(x)$, $b(x)$ and  $U(x)$ as:
\begin{eqnarray*}
  \mathcal M(x) &=& \left(U'(x)+b(x)\right)U(x)-\rmi\, U(x)a(x), \\
  \mathcal N(x) &=& a^2(x)-b^2(x)-\frac{\rmd}{\rmd x}\,U(x)b(x) \\
  && +\,\rmi\left(2a(x)b(x)+\frac{\rmd}{\rmd x}\,U(x)a(x)\right).
\end{eqnarray*}
\indent It should be noted that the defining assumption \eqref{1.1} can be generalized into $\widetilde\eta_+\mathcal H=\mathcal H^\dagger\widetilde\eta_+$. Applying Eqs.~\eqref{4}, \eqref{9} and \eqref{10} on both sides of the last equation and comparing their varying differential coefficients, we can recognized from the third-derivative that $b(x)=0$, which is a result already established in Sec.~II, while the second-derivative connects the potential to its complex-conjugate through:
\begin{equation}\label{11}
  V_+(x)=V_+^\ast(x)-4\rmi\, U(x)G'(x).
\end{equation}
\indent However, the coefficients corresponding to the first-derivative give the shape of the potential, where after integration, we get:
\begin{equation}\label{12}
  V_+(x)=F^2(x)-G^2(x)-\left[U(x)F(x)\right]'-2\rmi\, U(x)G'(x)+\epsilon,
\end{equation}
satisfying Eq.~\eqref{11} and $\epsilon$ is some real-constant of integration. The last remaining coefficient corresponds to the null-derivative and gives a pure-imaginary differential equation:
\begin{widetext}
\begin{eqnarray}\label{13}
  F^2(x)-[U(x)F(x)]'&=&\frac{G(x)}{G'(x)}\left(-F(x)F'(x)+\frac{1}{2}[U(x)F(x)]''\right)+\frac{1}{G'(x)}\bigg(\frac{1}{4}[U^2(x)G''(x)]'-\frac{G(x)}{4}[U(x)U''(x)]'\nonumber \\
  &+&\frac{U'(x)U(x)}{4}\bigg[\frac{G(x)}{U(x)}\bigg]''+\frac{U'^2(x)U(x)}{2}\bigg[\frac{G(x)}{U(x)}\bigg]'\bigg)-\frac{U''(x)U(x)}{4},
\end{eqnarray}
\end{widetext}
which is not easy to solve (here in terms of $F(x)$.)\\
\indent Obviously solving Eq.~\eqref{13} gives the generating function, i.e. $F(x)$ or $G(x)$, that identify $V_+(x)$ given by Eq.~\eqref{12}. To solve this equation we need use the $\widetilde\eta_+$-orthogonality condition, as deduced in Eq.~\eqref{2.10}, which suggests that any eigenfunction $\widetilde\Psi(x)\in\ker\widetilde\zeta^\dagger$ is related to $\mathcal H$ via $\widetilde\eta_+\Psi(x)\equiv \widetilde\zeta^\dagger\widetilde\Psi(x)=0$, with $\widetilde\Psi(x)=\widetilde\zeta\Psi(x)$.
\par Indeed let us define $\widetilde\Psi(x)$ as:
\[
\widetilde\Psi(x)=\left(\mathds 1-\widetilde\zeta\,
  \frac{1}{\widetilde\zeta^\dagger\widetilde\zeta}\,\widetilde\zeta^\dagger\right)\,\psi(x),
\]
for any $\psi(x)\in\ker\widetilde\zeta$. Then since $\widetilde\zeta^\dagger\,\widetilde\Psi\equiv\widetilde\zeta^\dagger\psi-\widetilde\zeta^\dagger\,\widetilde\zeta\left(\widetilde\zeta^\dagger\,\widetilde\zeta\right)^{-1}\widetilde\zeta^\dagger\,\psi = \widetilde\zeta^\dagger\psi-\widetilde\zeta^\dagger\psi=0$ we get $\widetilde\Psi\in\ker\widetilde\zeta\,^\dagger$, which leads 
to the following solution of the eigenfunction:
\begin{equation}\label{14}
\begin{split}
  \Psi(x) &= \rho(x)\psi(x) \\
          &= \exp\left[\rmi\int^x \rmd x'\frac{A(x')}{U(x')}\right]\psi(x)\\
          &\sim\exp\left[-\int^x \rmd x'\frac{F(x')}{U(x')}-\rmi\int^x \rmd x'\frac{G(x')-a(x')}{U(x')}\right],
\end{split}          
\end{equation}
up to normalization constant. The eigenfunction $\Psi(x)$ is then subjected to a gauge transformation of the kind: $\psi(x)\mapsto\Psi(x)=\rho(x)\psi(x)$, where the operator $\rho(x)=\sqrt{\widetilde\eta_+(x)}\equiv \exp\{-\rmi\,\alpha(x)/2\}$ implements a similarity transformation $\mathfrak h_{\rm Her}=\rho(x)\mathcal H\rho^{-1}(x)$ and $\alpha(x)$ is given by Eq.~\eqref{qq}. Here $\mathfrak h_{\rm Her}$ is called the {\it equivalent Hermitian counterpart Hamiltonian} analogue to the modified-Hamiltonian $\mathcal H$ (for more details, cf. Ref.~\cite{2}.)\\
\indent Indeed taking into account $\rho(x)$ and $\mathcal H$ as defined in Eqs.~\eqref{14} and \eqref{4}, respectively, the equivalent Hermitian counterpart Hamiltonian $\mathfrak h_{\rm Her}$ and its Hermitian self-adjoint are expressed as:
\begin{eqnarray}
  \mathfrak h_{\rm Her} &=&  -U^2(x)\frac{\rmd^2}{\rmd x^2}+2\left(2\rmi\, a(x)-U'(x)\right)U(x)\frac{\rmd}{\rmd x} \nonumber \\
  &+& 4a^2(x)+2\rmi \left[a(x)U(x)\right]'+V_+(x),\nonumber\\
  \mathfrak h^\dagger_{\rm Her}  &=& -U^2(x)\frac{\rmd^2}{\rmd x^2}+2\left(2\rmi\, a(x)-U'(x)\right)U(x)\frac{\rmd}{\rmd x} \nonumber \\
  &+& 4a^2(x)+2\rmi \left[a(x)U(x)\right]'+V_+^\ast(x),\nonumber
\end{eqnarray}
with $A(x)=a(x)\in\mathbb R$ and $b(x)=0$. Since $\mathfrak h_{\rm Her}$ is Hermitian, then $V_+(x)=V_+^\ast(x)$ and using Eqs.~\eqref{11} and \eqref{12}, keeping in mind that the mass-term $U(x)\neq0$, we deduce:
\begin{eqnarray*}
    G(x)&\equiv&\sqrt{\overline\epsilon}={\rm const.},\nonumber \\
    V_+(x)&=&F^2(x)-\left[U(x)F(x)\right]'\in\mathbb R,
\end{eqnarray*}
where, without loss of generality, we have set $\epsilon=\overline\epsilon$.\\
\indent Now using the associated Schr\"odinger equation $\mathcal H\Psi(x) = \mathcal E\Psi(x)$, with $\mathcal E = \mathcal E_{\rm  Re} + \rmi\,\mathcal E_{\rm  Im}$, one obtain the differential equation:
\begin{eqnarray}\label{15}
  -\mathcal E_{\rm Im} + \rmi (\mathcal E_{\rm  Re}-\epsilon)&=& 2F(x)G(x)+U(x)G'(x) \nonumber \\ && -U'(x)G(x),
\end{eqnarray}
where $\epsilon$ is a constant already deduced in Eq.~\eqref{12}. In order to solve suitably Eq.~\eqref{15} we notice that the right-hand side is a
function while the left-hand side is a constant; thus we can assume that the both sides are equal to some constant, considered here as null, which requires that $\mathcal E_{\rm  Re}=\epsilon$ and $\mathcal E_{\rm  Im}=0$. Hence, as it was expected in Eq.~\eqref{2.10}, the energy eigenvalues are {\it real}.\\
\indent In these settings, using Eq.~\eqref{15}, we end up by relating $F(x)$ to $G(x)$ and $U(x)$ through the differential equation:
\begin{eqnarray}\label{16}
  F(x) = \frac{G(x)}{2}\bigg[\frac{U(x)}{G(x)}\bigg]',
\end{eqnarray}
and the interested reader can check that Eq.~\eqref{16} is the solution of Eq.~\eqref{13}. Therefore we can interpret $F(x)$, as well as $G(x)$, as the generating function leading to identify the potential $V_+(x)$ in Eq.~\eqref{12}.\\
\indent We now turn to the weak pseudo-Hermiticity operator by introducing the first-order differential realization for $\eta_-$. In this case $\eta_-$ is anti-Hermitian which amounts to writing $\eta_-^\dagger=-\eta_-$; therefore the modified-Hamiltonian $\mathcal H$ can be relaxed to be weak pseudo-Hermitian, such that:
\begin{equation}\label{17}
\begin{split}
  \eta_- &= U(x)\frac{\rmd}{\rmd x}+w(x), \\
  \eta_-^\dagger &=-\frac{\rmd}{\rmd x}\,U(x)+w^\ast(x),
  \end{split}
\end{equation}
with $w(x)=f(x)+\rmi\, g(x)$, and $f(x)$ and $g(x)$ are an arbitrary real functions. Using the same shift on the momentum as above, it turns out that:
\begin{equation}\label{18}
\begin{split}
  \eta_-&\quad\mapsto\quad\widetilde\eta_- = U(x)\frac{\rmd}{\rmd x}-\rmi A(x)+w(x), \\  \eta_-^\dagger&\quad\mapsto\quad\widetilde\eta_-^\dagger = -\frac{\rmd}{\rmd x}\,U(x)+\rmi A^\ast(x)+w^\ast(x),
\end{split}
\end{equation}
and the anti-Hermitian condition $\widetilde\eta_-^\dagger=-\widetilde\eta_-$ leads to the relation:
\begin{eqnarray}\label{19}
  f(x)+b(x)=\frac{U'(x)}{2}.
\end{eqnarray}
\indent Letting both sides of $\widetilde\eta_-\mathcal H = \mathcal H^\dagger\widetilde\eta_-$ act on every function and comparing their
varying differential coefficients, one deduce from the second-derivative, once more, that $b(x)=0$. Thus, from Eq.~\eqref{19}, the generating
function $f(x)$ becomes:
\begin{eqnarray}\label{20}
  f(x) = \frac{U'(x)}{2},
\end{eqnarray}
while the null-derivative (by simple integration) and the first-derivative terms give the real and imaginary parts of the potential $V_-(x)$, respectively, as:
\begin{eqnarray*}
  V^{(\rm Re)}_-(x) &=& (f+ig)^2-w(x)U'(x)-\frac{1}{2}\,U(x)U''(x)+\gamma, \\
  V_-^{(\rm Im)}(x) &=& \rmi\, U(x)w'(x)-\frac{\rmi}{2}U(x)U''(x),
\end{eqnarray*}
where $\gamma$ is a constant of integration. Combining last two results, taking into account Eq.~\eqref{20}, we obtain:
\begin{eqnarray}\label{21}
  V_-(x) &=& (f+ig)^2-\frac{d}{dx}U(x)(f+ig)+\gamma \nonumber \\
             &=& w^2(x)-\frac{d}{dx}\,U(x)w(x)+\gamma.
\end{eqnarray}

\section{Complementarity and coordinate transformation: connection via similarity transformations}%

\noindent The arguments developed in the preceding two sections led us in a natural way to investigate the connection between the concept of
complementarity and the notion of coordinate transformations through their respective generating functions $F(x)$ and $f(x)$, and also to investigate if there exists a relationship connecting both $\widetilde\eta_+$ and $\widetilde\eta_-$. The main idea of such an achievement lies in the fact that both generating functions belong to the same space configuration, $\{\bf{\mathfrak X}\}$, and this lets us think that there should be a coordinate transformation connecting them.
\subsection{Complementarity vs. coordinate transformations}%
\noindent In mathematical terms we admit the existence of some coordinate transformations, say $x\equiv x(\xi)$, that changes $F(x)$ into $f(\xi)$ in the following way:
\begin{equation}\label{xx}
  F(x) = \frac{G(x)}{2}\bigg[\frac{U(x)}{G(x)}\bigg]'
  \quad\xrightarrow{x\equiv x(\xi)} \quad  f(\xi) = \frac{\cal U'(\xi)}{2}.
\end{equation}
\indent An interesting way to solve this problem is to build a differential equation from Eq.~\eqref{16} and assume that it is maintained invariant under the action of a coordinate transformation. In fact, Eq.~\eqref{16} can be recast as:
\begin{eqnarray}\label{22}
  U(x)\frac{\rmd Z(x)}{\rmd x} = 2F(x)Z(x),
\end{eqnarray}
where $Z(x)=U(x)/G(x)$ with $G(x)\neq0$.
\par In particular the introduction of two new functions, $R(\xi)$ and $S(\xi)$, with the help of a coordinate transformation $x=x(\xi)$, change the functions $U(x),\, F(x)$ and $Z(x)$, respectively, in the following way (cf. Eqs.~(3.9) in Ref. \cite{21}):
\begin{eqnarray}
  U(x) \quad&\rightarrow&\quad \mathcal U[\xi(x)] = U(x)\, \frac{\rmd\xi(x)}{\rmd x},         \label{23} \\
  F(x) \quad&\rightarrow&\quad \mathcal F[\xi(x)] = F(x)\, S(\xi),                            \label{24} \\
  Z(x) \quad&\rightarrow&\quad \mathcal Z[\xi(x)] = Z(x)\, R(\xi).                            \label{25}
\end{eqnarray}
\indent Using Eqs.~\eqref{23}-\eqref{25}, the similar differential equation to Eq.~\eqref{22} can be obtained for the function $\mathcal Z(\xi)$,
i.e.
\begin{eqnarray}\label{26}
  \mathcal U(\xi)\frac{\rmd \mathcal Z(\xi)}{\rmd \xi} = 2\, \mathcal F(\xi) \mathcal Z(\xi),
\end{eqnarray}
with the following restriction on the functions $R(\xi)$ and $S(\xi)$, namely:
\[
R(\xi)S(\xi)\equiv 1\qquad\Rightarrow\qquad R(\xi)=S^{-1}(\xi).
\]
\indent Substituting once more Eqs.~\eqref{23}-\eqref{25} into Eq.~\eqref{26} we are then led to the differential equation:
\begin{equation}\label{27}
  U(x)\frac{\rmd Z(x)}{\rmd x} = 2\left(S(x)F(x)-U(x)\frac{\rmd}{\rmd x}\ln\sqrt{R(x)}\right)Z(x),
\end{equation}
and by identifying to Eq.~\eqref{22}, using the identity $R(\xi)=S^{-1}(\xi)$, we end up with a differential equation in terms of $R(x)$:
\begin{equation*}
F(x)=S(x)F(x)-U(x)\frac{\rmd}{\rmd x}\ln\sqrt{R(x)},
\end{equation*}
i.e.
\begin{eqnarray}\label{4.6}
    F(x)=\frac{U(x)R(x)}{1-R(x)}\frac{\rmd}{\rmd x}\ln\sqrt{R(x)},
\end{eqnarray}
whose its solution is given by:
\begin{equation}\label{4.7}
    R(x)=1+\delta\exp\left\{-2\int^x\frac{F(x')}{U(x')}\,\rmd x'\right\},
\end{equation}
and so:
\[
S(x)=\left(1+\delta\exp\left\{-2\int^x\frac{F(x')}{U(x')}\,\rmd x'\right\}\right)^{-1},
\]
with $\delta$ is some constant of integration. With these settings we obtain transformation which can be regarded as a {\it similarity transformation} relating $F(x)$ to $f(x)$ through the mapping (cf, for instance, the similarity with Eqs.~(3.11) in Ref.~\cite{21}):
\begin{eqnarray}\label{28}
  F(x)\,\mapsto \, f(x)&\equiv& S(x)F(x)\nonumber \\
   &=&F(x) + U(x)\frac{\rmd}{\rmd x}\ln\sqrt{R(x)}.
\end{eqnarray}
\indent Let us redefine the coordinate transformation on $F(x)$ following Eq.~\eqref{24} as:
\begin{equation}\label{29}
  F(x) \,\rightarrow\, \mathcal F(\xi) = F[x(\xi)]S(\xi) \equiv F[x(\xi)]\frac{\rmd\xi(x)}{\rmd x},
\end{equation}
by defining $S[\xi(x)]=\rmd\xi(x)/\rmd x$,
{\rm i.e.} $(S^{-1}[x(\xi)]\equiv R[x(\xi)]=\rmd x(\xi)/\rmd\xi)$, and from Eqs.~\eqref{16} and \eqref{28} we get the identity:
\begin{equation}\label{30}
  f(\xi) = F[x(\xi)]S(\xi) = \frac{G[x(\xi)]S(\xi)}{2}\left(\frac{U[x(\xi)]}{G[x(\xi)]}\right)'.
\end{equation}
\indent Since we are dealing with two unknown functions $R(\xi)$ and $S(\xi)$ then the best we can do is to fix one of them and connect it either to $F(x)$ or $G(x)$. It is here that we invoke a restriction on Eq.~\eqref{30} such that
$G[x(\xi)]S(\xi)=1\equiv R(\xi) S(\xi)$, which leads to define the generating function $G(x)$, using Eq.~\eqref{29}, as:
\begin{eqnarray}\label{31}
  G[x(\xi)]\equiv S^{-1}(\xi)=\frac{\rmd x(\xi)}{\rmd \xi}=R(\xi),
\end{eqnarray}
and taking into account both Eqs.~\eqref{23} and \eqref{31}, then Eq.~\eqref{30} can be simplified to:
\begin{eqnarray}\label{32}
  f[\xi(x)] &=& \frac{1}{2}\bigg(U[x(\xi)]\frac{\rmd\xi(x)}{\rmd x}\bigg)' \nonumber \\
            &=& \frac{\mathcal U'[\xi(x)]}{2},
\end{eqnarray}
which completes the proof of our assertion given in Eq.~\eqref{xx}.
\subsection{($\widetilde\eta_+,\widetilde\eta_-$)-connection vs. similarity transformation}%
\noindent We pursue the point addressed above further by constructing the relation that gives rise to the identity connecting both $\widetilde\eta_+$ and $\widetilde\eta_-$ which, to this order, is given by the similarity transformation $\underline{\bf D} = \mathcal B(x)\,\underline{\bf d}\,\mathcal B^{-1}(x)$, where $\underline{\bf D}$ and $\underline{\bf d}$ are some operators to be determined and $\mathcal B(x)$ is a function that implements the similarity transformation given in Eq.~\eqref{28}. However, the requirement that the only knowledge for $\widetilde\eta_+$ (resp. $\widetilde\eta_-$) is its Hermiticity (resp. anti-Hermiticity) suggests that the similarity transformation is {\it not unique}, further that $\widetilde\eta_+$ is a second-order differential operator and $\widetilde\eta_-$ is of the first-order.\\
\indent To cope with this difficulty, it is worth mentioning the existence of an underlying first-order differential operator $\widetilde\zeta$ introduced in Eq.~\eqref{8}, related to $\widetilde\eta_+\equiv \widetilde\zeta^\dagger \widetilde\zeta$, which is better adapted to different calculations rather than $\widetilde\eta_+$. Thus, the implementation of the similarity transformation mentioned above concerns only the two operators $\underline{\bf D}=\widetilde\zeta$ and $\underline{\bf d}=\widetilde\eta_-$, and also the real-function $\mathcal B(x)=R^{1/2}(x)$, such that:
\begin{equation}\label{4.15}
\begin{split}
 \widetilde\zeta &= R^{1/2}(x) \widetilde\eta_- R^{-1/2}(x),\\
 \widetilde\zeta\,^\dagger &= R^{-1/2}(x)\widetilde\eta_-^\dagger R^{1/2}(x).
\end{split}
\end{equation}
\indent Inserting Eq.~\eqref{18} into Eq.~\eqref{4.15}, taking into account $A(x)=a(x)\in\mathbb R$ and $b(x)=0$ constraints already obtained in Secs.~II and III, and identifying with Eq.~\eqref{28}, we get $G(x)=g(x)$. Thus we can say that the direct consequence of the similarity transformation \eqref{4.15} is that the real parts of $W(x)\left(=F(x)+\rmi\, G(x)\right)$ and $w(x)\left(=f(x)+\rmi\, g(x)\right)$ are connected to each other through a coordinate transformations \eqref{xx}, while the imaginary parts remain the same.\\
\indent Thus acting $\widetilde\zeta^\dagger$ on $\widetilde\zeta$, we get the relationship which we looking for, i.e.
\begin{eqnarray*}
  \widetilde\eta_+ &\equiv& \widetilde\zeta^\dagger \widetilde\zeta \nonumber \\
    &=& R^{-1/2}(x)\,\,\widetilde\eta_-^\dagger\, R(x)\,\widetilde\eta_-\,R^{-1/2}(x),
\end{eqnarray*}
connecting $\widetilde\eta_+$ to $\widetilde\eta_-$ by means of $R(x)$. Doing some little algebra shows that the operator $\widetilde\eta_+$ can be written either in its {\it factorizable} form as a product of a pair of first-order differential operators:
\begin{equation}\label{4.16}
  \widetilde\eta_+ = \left(\widetilde\eta^\dagger_--\frac{U(x)}{2}\frac{\rmd}{\rmd x}\ln R(x)\right)\left(\widetilde\eta_--\frac{U(x)}{2}\frac{\rmd}{\rmd x}\ln R(x)\right),
\end{equation}
as well as in its {\it decomposition} form:
\begin{eqnarray}\label{4.17}
  \widetilde\eta_+ &=& \widetilde\eta_-^\dagger\widetilde\eta_-
                     +\frac{1}{2}\frac{U^2(x) R''(x)}{R(x)}+\frac{U(x)U'(x)R'(x)}{R(x)}\nonumber \\
                     &&-\frac{R(x)-3}{R(x)-1}\left(\frac{U(x) R'(x)}{2R(x)}\right)^2,
\end{eqnarray}
where we used Eqs.~\eqref{4.6} and \eqref{28} to deduce Eq.~\eqref{4.17}.\\
\indent May be the main question arises now is what kind of coordinate transformations we are looking for or, otherwise, should we avoid?
Answering this question comes to take full advantage on the expressions under a square root and in a denominator given by Eq.~\eqref{4.16} and Eq.~\eqref{4.17}, respectively, in the sense that the both expressions should be free from singularities, i.e. $R(x)\neq 0$ and $R(x)\neq 1$. Indeed these facts suggest, using Eq.~\eqref{31}, that if we consider that $R(\xi) \equiv \rmd x(\xi)/\rmd\xi=0$ and $R(\xi) \equiv \rmd x(\xi)/\rmd\xi=1$, then it is easy to convince ourselves that the specific choice of coordinate transformations can not be a {\it constant transformation} $x(\xi)=k_1$, no longer a {\it linear coordinate transformation} $x(\xi)=k_0\xi+k_2$, with $k_0=1$ and $k_{1,2}$ are constants. Then we can conclude that the kind of coordinate transformations which we are looking for are both of linear (with $k_0\neq 1$) and nonlinear types.\\
\indent If we set $\widetilde\eta_+\equiv \widetilde\zeta^\dagger \widetilde\zeta=\mathds 1$ in the pseudo-Hermiticity condition $\widetilde\eta_+\mathcal H=\mathcal H^\dagger\widetilde\eta_+$, then the modified-Hamiltonian $\mathcal H$ is Hermitian. It is apparent in this case that $\widetilde\zeta$ is an unitary operator, i.e. $\widetilde\zeta^\dagger=\widetilde\zeta^{-1}$.
\subsection{Case of constant mass, mapping $\eta_+ = \eta_-^\dagger \eta_-$ and B\"acklund transformations}%
\noindent Before discussing the possibility of considering $\eta_+$ as a product of $\eta_-$ and its hermitian-conjugate, i.e. $\eta_+=\eta_-^\dagger \eta_-$, associated to the case of a constant mass (CM), let us first look at the general case corresponding to a PDM system given by the factorization $\widetilde\eta_+ = \widetilde\eta_-^\dagger\widetilde\eta_-$.\\
\indent This factorization can be given if, and only if, the last three terms of Eq.~\eqref{4.17} are null, which yields an ordinary differential equation:
\begin{equation}\label{4.18}
  \frac{\rmd^2 R(x)}{\rmd x^2} = -\theta(x)\frac{\rmd R(x)}{\rmd x}+\chi[R(x)]\left(\frac{\rmd R(x)}{\rmd x}\right)^2,
\end{equation}
where the functions $\theta(x)$ and $\chi(R)$ are defined by:
\[
\theta(x)=2\frac{\rmd}{\rmd x}\ln U(x),\quad\chi[R(x)]=\frac{R(x)-3}{2R(x)(R(x)-1)}.
\]
\indent Dividing Eq.~\eqref{4.18} by $R'(x)$ leads to a total differential equation with separation of variables, whose its solution,
\begin{eqnarray}
  R'(x) = \frac{\lambda_1}{U^2(x)}\frac{R^{3/2}(x)}{R(x)-1},\nonumber
\end{eqnarray}
is a first-order differential equation easy to solve. Indeed integrating this latter by separation of variables, squaring and solving the resulting equation in $R(x)$, we obtain, for a PDM system, the following solution:
\begin{equation}\label{4.19}
  R_{\rm PDM}(x) = \frac{\sigma^2(x)}{8}-1\pm\frac{\sigma(x)}{2}\sqrt{\frac{\sigma^2(x)}{16}-1},
\end{equation}
with,
\begin{eqnarray}\label{4.20}
  \sigma(x) = \lambda_1\int^x \frac{\rmd x'}{U^2(x')}+\lambda_2,
\end{eqnarray}
where $\lambda_{1,2}$ are some constants of integration. Furthermore, by using Eq.~\eqref{31}, the specific choice of coordinate transformations can be readily recognized through the integral:
\begin{equation}\label{4.21}
  \xi_{\rm PDM}(x) = c+\int^x\frac{\rmd x'}{\frac{\sigma^2(x')}{8}-1\pm\frac{\sigma(x')}{2}\sqrt{\frac{\sigma^2(x')}{16}-1}},
\end{equation}
where $c$ is a constant of integration.\\
\indent Obviously the case of a constant mass (CM) can be recovered from a PDM case by removing a shift on the momentum introduced in Sec.~II and imposing the constraint $U(x)\equiv1/\sqrt{2m(x)}=1.$ This latter reduces the original differential equation \eqref{4.18}, by means of substitution $\theta(x)=0$, to:
\begin{eqnarray}\label{4.22}
  \frac{\rmd^2 R(x)}{\rmd x^2} = \chi[R(x)]\,\varphi\left(\frac{\rmd R(x)}{\rmd x}\right),
\end{eqnarray}
with $\varphi\left(R'_x\right)=R'^2(x)$, whose its solution is given by:
\[
  R_{\rm CM}(x) = \frac{x^2}{8}-1\pm\frac{x}{2}\sqrt{\frac{x^2}{16}-1},
\]
where, without loss of generality, we set $\lambda_1=1$ and $\lambda_2=0$, in which a PDM solutions given in Eqs.~\eqref{4.20} and \eqref{4.21} are reduced to the specific coordinate transformations featuring a CM system:
\begin{eqnarray}\label{4.23}
  \xi_{\rm CM}(x) = \int^x\frac{\rmd x'}{\frac{x'^2}{8}-1\pm\frac{x'}{2}\sqrt{\frac{x'^2}{16}-1}},
\end{eqnarray}
with $c=0$. However it is worth mentioning that the use of Eq.~\eqref{23}, keeping in mind that $U(x)=1$, allows us to deduce the inverse transform of Eq.~\eqref{4.23},
\begin{equation*}
  x_{\rm CM}(\xi)=\int^\xi\frac{\rmd\xi'}{\mathcal U(\xi')},
\end{equation*}
leading to identify the modified-mass term, $\mathcal U(\xi)\neq1$, as:
\[
\mathcal U[\xi_{\rm CM}(x)] = \left(\frac{x^2}{8}-1\pm\frac{x}{2}\sqrt{\frac{x^2}{16}-1}\,\right)^{-1}.
\]
\indent Moreover the differential equation \eqref{4.22}, admitting an exact solution, is subjected to some transformations, which enable us to understand in what circumstances the special factorization $\eta_+=\eta_-^\dagger\eta_-$ can arises. To this end we observe two kind of transformations labeled by $\widehat{\mathbb S}$ and $\widehat{\mathbb B}$.
\subsubsection{$\widehat{\mathbb S}$-transformations.}
\noindent We first make the transformation, denoted here $\widehat{\mathbb S}$, by assuming $R$ as an independent variable, while $x$ becomes the dependent one. Then Eq.~\eqref{4.22} is transformed into a differential equation of the similar form for $x\equiv x\left(R\right)$:
\begin{eqnarray}\label{4.24}
  \widehat{\mathbb S}:\quad \frac{\rmd^2 x(R)}{\rmd R^2} = \widetilde\chi(x)\varphi^\sharp\left(\frac{\rmd x(R)}{\rmd R}\right),
\end{eqnarray}
where $\widetilde\chi(x) \equiv \widetilde\chi[x(R)] = \chi(R)$, and $\varphi^\sharp\left(q\right)=-q^3\varphi(1/q)$ with $q \equiv x'_{R}= \rmd x(R)/\rmd R$.
\subsubsection{$\widehat{\mathbb B}$-transformations.}
\noindent The second type of transformations are those known as a B\"acklund transformations (cf. Ref.~\cite{22} for more detailed discussions) and denoted hereinafter by $\widehat{\mathbb B}$.
\par According to the classical definition, a B\"acklund transformation between a $(n+1)$-ordinary differential equations (ODEs), i.e.
\begin{equation*}
\begin{split}
E(R,x)&=0, \\
\overline E(\overline R,\overline x)&=0, \\
\overline{\overline E}\left(\overline{\overline R},\overline{\overline x}\right)&=0, \\
&\vdots \\
\overline E^{(n)}\left(\overline R^{(n)},\overline x^{(n)}\right)&=0,
\end{split}
\end{equation*}
where $n$ (i.e. the process) is finite, is a pair of relationship:
\[
\mathfrak R_{ij}\left(\overline R^{(i)},\overline x^{(i)}\Big|\overline R^{(j)},\overline x^{(j)}\right)=0,\quad (0\leq i<j \leq n),
\]
with some additional identities connecting the old-$(R,x)$ variables to the new-$(\overline R^{(i)},\overline x^{(i)})$ variables in order to get the ODEs $E$ and $\overline E^{(i)}$. Therefore if the solution of one of them is known, say $\overline E$, then we can obtain all the solutions of $E$ and $\overline E^{(i)}$, where $i=2,3,\ldots,n$. Unfortunately, there is no known systematic approach for providing a B\"acklund transformations, it is usually not easy to get them as we will see below.
\par Indeed we display here the explicit form of a B\"acklund transformations leading to take a special factorization $\eta_+ = \eta_-^\dagger \eta_-$ into consideration. To this end we start by introducing the following B\"acklund transformation to Eq.~\eqref{4.22}:
\begin{equation}\label{4.25}
\begin{split}
  \widehat{\mathbb B}:\quad\overline x &= \int\frac{\rmd p}{\varphi(p)},\\
  \overline{R} &= \int \lambda_1\rmd x,
\end{split}
\end{equation}
where $p\equiv \rmd R/\rmd x=R'(x)=1/q$ and $\lambda_1$ is a constant, leads to an equation of the similar form for the new function $\overline R\equiv \overline R(\overline x)$, i.e.
\begin{equation*}
  \frac{\rmd^2\overline R}{\rmd\overline x\,^2}\equiv\overline\chi\left(\overline R\right)
              \overline\varphi\left(\overline p\right) = -\lambda_1\,\frac{R'(x)}{\chi^3(R)}\frac{\rmd\chi(R)}{\rmd R},
\end{equation*}
where $\overline p\equiv \rmd\overline R/\rmd \overline x=\overline R\,'({\overline x})$ and $\lambda_1=\overline p\,\chi(R)$. The new functions $\overline\chi\left(\overline R\right)$ and $\overline\varphi\left(\overline p\right)$ are given through the relationship:
\begin{equation}\label{4.26}
\begin{split}
  \overline\chi\left(\overline R\right)   &= R'(x) = p, \\
  \overline\varphi\left(\overline p\right)&=-\frac{\lambda_1}{\chi^3(R)}\frac{\rmd\chi(R)}{\rmd R}.
\end{split}  
\end{equation}
\indent Now the two-fold application of $\hat{\mathbb B}$, i.e. $n=2$, given by Eq.~\eqref{4.25}, using Eq.~\eqref{4.26},
\begin{equation}\label{yy}
\begin{split}
  \widehat{\mathbb B}:\quad
  \overline{\overline x} &= \int\frac{\rmd\overline p}{\overline\varphi\left(\overline p\right)}=\int R'(x) \chi(R)\rmd x, \\
  \overline{\overline R} &= \int \lambda_2\, \rmd\overline x = \int \overline{\overline p}\,\frac{p\,\rmd p}{\varphi(p)},
\end{split}
\end{equation}
with $\lambda_2=\overline{\overline p}\,\overline\chi\left(\overline R\right)$, transforms once more the original equation into:
\begin{eqnarray*}
  \frac{\rmd^2\overline{\overline R}}{\rmd\overline{\overline x}\,^2}
        &\equiv &
        \overline{\overline\chi}\left(\overline{\overline R}\right)
        \overline{\overline\varphi}\left(\overline{\overline p}\right) \\
        &=& -\lambda_2\frac{\varphi(p)}{R'^3(x)},
\end{eqnarray*}
with $\overline{\overline p}\equiv \rmd\overline{\overline R}/\rmd\overline{\overline x}=\overline{\overline R}\,'({\overline{\overline x}})$ and thus the correspondence connecting $(\chi,\varphi)$ to $\left(\overline{\overline\chi},\overline{\overline\varphi}\right)$, respectively, are given by:
\begin{equation}\label{4.27}
\begin{split}
   \overline{\overline\chi}\left(\overline{\overline R}\right) &= \frac{\lambda_2}{R'(x)} \equiv \frac{\lambda_2}{p}, \\
   \overline{\overline\varphi}\left(\overline{\overline p}\right) &= -\frac{\varphi\left(R'_x\right)}{R'^2(x)}
          \equiv -\frac{\varphi\left(p\right)}{p^2}.
\end{split}
\end{equation}
\indent Finally the three-fold application of $\hat{\mathbb B}$, i.e. $n=3$, using Eqs.~\eqref{4.25}-\eqref{4.27}, yields the original differential equation Eq.~\eqref{4.22}, i.e.
\begin{widetext}
\begin{equation}\label{zz}
\begin{split}
  \widehat{\mathbb B}:\quad
  \overline{\overline{\overline x}} &=
           \int\frac{\rmd \overline{\overline p}}{\overline{\overline\varphi}\left(\overline{\overline p}\right)}
           =\int \lambda_2\,\chi(R)\,\rmd x
           =\int \overline{\overline p}\,R'(x)\chi(R)\,\rmd x \equiv x,\\
  \overline{\overline{\overline R}} &= \int \lambda_3\,\rmd\overline{\overline x}
           = \int \overline{\overline{\overline p}}\,\overline{\overline\chi}\left(\overline{\overline R}\right)R'(x)\chi(R)\,\rmd x
           = \int \overline{\overline{\overline p}}\,\frac{\lambda_2}{R'(x)}\chi(R)\,\rmd R \equiv R,
\end{split}
\end{equation}
\end{widetext}
with $\lambda_3=\overline{\overline{\overline p}}\,\overline{\overline\chi}\left(\overline{\overline R}\right)$, if and only if the following identities are satisfied:
\begin{eqnarray}\label{4.28}
\overline{\overline p}\,R'(x)\chi(R) = 1
\quad {\rm and} \quad
\lambda_2\,\overline{\overline{\overline p}}\,\chi(R) = R'(x).
\end{eqnarray}
\par Since $\lambda_2=\overline{\overline p}\,\overline\chi\left(\overline R\right)$, we get from Eq.~\eqref{4.28} the relations satisfying both of $\overline{\overline p}$ and $\overline{\overline{\overline p}}$, i.e.
\begin{eqnarray}\label{4.29}
  \overline{\overline p} = \frac{1}{R'(x)\chi(R)}
  \qquad {\rm and} \qquad
  \overline{\overline{\overline p}} \equiv p = R'(x),
\end{eqnarray}
since indeed the simple calculation gives the original differential equation \eqref{4.22}:
\begin{eqnarray*}
\frac{\rmd^2 \overline{\overline{\overline R}}}{\rmd \overline{\overline{\overline x}}\,^2}
= \frac{\rmd x}{\rmd \overline{\overline{\overline x}}}\cdot\frac{\rmd}{\rmd x}\left(\frac{\rmd\overline{\overline{\overline R}}}{\rmd \overline{\overline{\overline x}}}\right)
= \frac{\rmd}{\rmd x}\,\overline{\overline{\overline p}} = \frac{\rmd}{\rmd x}\,p &\equiv & \frac{\rmd^2 R}{\rmd x^2} \nonumber \\
&=& \chi(R)\varphi\left(\frac{\rmd R}{\rmd x}\right),
\end{eqnarray*}
and so,
\[
\overline{\overline{\overline {\chi}}}\left(\overline{\overline{\overline R}}\right)\equiv \chi(R)
\qquad{\rm and}\qquad
\overline{\overline{\overline {\varphi}}}\left(\frac{\rmd\overline{\overline{\overline R}}}{{\rmd\overline{\overline{\overline x}}}}\right)
\equiv \varphi\left(\frac{\rmd R}{\rmd x}\right),
\]
keeping in mind that $R(x)$ is a solution of ODE \eqref{4.22}.

\begin{figure}[h]
\begin{xy}
\xymatrixrowsep{1.3cm}
\xymatrixcolsep{1.2cm}
\xymatrix{
{\big\{\chi,\varphi\big\}}\ar@{--}[r]&\ar@{--}[r]^{\widehat{\mathbb S}}&\ar@{-->}[r]&\big\{\chi,\varphi^\sharp\big\} \\
&{\left\{\overline{\overline\chi},\overline{\overline\varphi}\right\}}\ar@{->}[lu]^{\widehat{\mathbb B}}\ar@{-->}[r]^{\widehat{\mathbb S}}&
               \left\{\overline{\overline\chi},\overline{\overline\varphi}\,^\sharp\right\}\ar@{->}[ur]^{\widehat{\mathbb B}}  \\
               \big\{\overline\chi,\overline\varphi\big\}\ar@{->}[ur]^{\widehat{\mathbb B}}\ar@{<-}[uu]^{\widehat{\mathbb B}}
  \ar@{--}[r]&\ar@{--}[r]^{\widehat{\mathbb S}}&\ar@{-->}[r]&\big\{\overline\chi,\overline\varphi\,^\sharp\big\}\ar@{<-}[uu]^{\widehat{\mathbb B}}
  \ar@{->}[lu]^{\widehat{\mathbb B}} \\
}
\end{xy}
\end{figure}
\par Therefore the ODE \eqref{4.22}, giving the special factorization $\eta_+=\eta_-^\dagger\eta_-$, can be considered as an (invariant) {\it auto-B\"acklund transformations} after the three-fold application of $\widehat{\mathbb B}$, ($\widehat{\mathbb B}^3=\mathds 1$, i.e. $\widehat{\mathbb B}^2=\widehat{\mathbb B}^{-1}$), which means that Eq.~\eqref{yy} and the inverse of Eq.~\eqref{4.25} are equivalent. However the correspondence \eqref{4.25} and \eqref{yy} are the well-known {\it hetero-B\"acklund transformations}. Obviously the ODE deduced from the $\widehat{\mathbb S}$-transformation, Eq.~\eqref{4.24}, is invariant under the same B\"acklund transformations, Eqs.~ \eqref{4.25}, \eqref{yy} and \eqref{zz}, as summarized in the diagram below, where, for convenience we have used in the diagram the symbolic notation $\left\{\chi,\varphi\right\}$ (introduced in Ref.~\cite{23}, cf. subsection 2.7.6.) to denote the differential equation \eqref{4.22}.
\par Hence identifying the mapping $\eta_+=\eta^\dagger_-\eta_-$ has a direct consequence of a consecutive application of the similar transformation $\widehat{\mathbb S}$ (dashed arrows) and a B\"acklund transformations $\widehat{\mathbb B}$ (solid arrows) is the generation of six different ordinary differential equations of the form of Eq.~\eqref{4.22}, where as shown in the diagram $\widehat{\mathbb B}^3=\mathds 1$ and $\widehat{\mathbb S}^{-1}\circ\widehat{\mathbb B}\circ\widehat{\mathbb S}=\widehat{\mathbb B}$, and conversely $\widehat{\mathbb B}^{-1}\circ\widehat{\mathbb S}\circ\widehat{\mathbb B}=\widehat{\mathbb S}$, since both transformations commute.

\section*{Conclusion}%

In this paper, we have tackle the concept of complementarity in a new way and completely different from that exposed in Mostafazadeh's paper \cite{4}. We have given an explicit proof, in the framework of PDM, that pseudo-Hermiticity and weak pseudo-Hermiticity are essentially complementary to one another. Accordingly, the statement that made in the pioneering work \cite{7} about "the complementarity between pseudo-Hermiticity and weak pseudo-Hermiticity" can indeed be interpreted as coordinate transformations connecting their respective generating functions $f(x)$ and $F(x)$. This connection led us to construct a similarity transformation, implemented by the function $R(x)$, connecting both $\widetilde\eta_+$ and $\widetilde\eta_-$. We also asked our results what kind of coordinate transformations could reproduce the connection between $\widetilde\eta_+$ and $\widetilde\eta_-$ and it turns out that they must satisfy both of linear (with $k_0\neq 1$) and nonlinear transformations.
\par Furthermore, the case of a constant mass is discussed and we have shown that a special factorization $\eta_+=\eta_-^\dagger\eta_-$ can only be given if one specifies some features of an ordinary differential equation. We have shown that those features are subjected to some transformations; among them the similar and B\"acklund transformations have been constructed.

\end{document}